# Mixed Convective Heat Transfer Enhancement in a Ventilated Cavity by Flow Modulation via Rotating Plate

Aminul Islam[1], Monoranjan Debnath Rony[1], Mahmudul Islam[1], Emdadul Haque Chowdhury[1]
Mohammad Nasim Hasan[1,*]

[1]Department of Mechanical Engineering, Bangladesh University of Engineering and Technology, Dhaka-1000, Bangladesh.

[*]Corresponding author *Email address: nasim@me.buet.ac.bd*

**Abstract**

The present study numerically explores the mixed convection phenomena in a differentially heated ventilated square cavity with active flow modulation via a rotating plate. Forced convection flow in the cavity is attained by maintaining an external fluid flow through an opening at the bottom of the left cavity wall while leaving it through another opening at the right cavity wall. A counter-clockwise rotating plate at the center of the cavity acts as active flow modulator. Moving mesh approach is used for the rotation of the plate and the numerical solution is achieved using Arbitrary Lagrangian-Eulerian (ALE) finite element formulation with a quadrilateral discretization scheme. Transient parametric simulations have been performed for various frequency of the rotating plate for a fixed Reynolds number ($Re$) of 100 based on maximum inlet flow velocity while the Richardson number ($Ri$) is maintained at unity. Heat transfer performance has been evaluated in terms of spatially averaged Nusselt number and time-averaged Nusselt number along the heated wall. Power spectrum analysis in the frequency domain obtained from the fast Fourier transform (FFT) analysis indicates that thermal frequency and plate frequency start to deviate from each other at higher values of velocity ratio ($> 4$).

## 1. Introduction

Due to the widespread importance in numerous important engineering and industrial applications, the transport process involved in a mixed convection heat transfer inside a ventilated cavity is a matter of great interest. The addition of one or more rotating elements into the cavity can intensify this kind of transport phenomenon. Therefore, the buoyancy driven flow generated from the temperature gradient of differentially heated side walls is coupled with the forced flow developed by fluid flow into the ventilated cavity together with the centrifugal action of the rotating plates. The complex flow and thermal transport behavior evolved inside such an enclosure can be effectively realized with the aid of computational fluid dynamics. This type of cavity has many well-known applications, such as Food Processing Technologies [1], crystal formation, high performance building insulation, Solar Distiller [2], solar energy collectors, nuclear reactors, electronic device cooling [3], chemical processing equipments [4], drying technology, float glass processing [5], chemical mixing devices [6] etc.

Mixed convection is generally characterized by a dimensionless parameter called Richardson number ($Ri$) within the range of $0.1 \leq Ri \leq 10$. Richardson number (Ri) indicates the relative significance between forced convection and natural convection, and mathematically expressed by the ratio between buoyancy force and inertia force. If $Ri \gg 1$, the governing mechanism for heat transfer is the buoyancy effect, whereas, $Ri \ll 1$ corresponds to the prominence of forced convection. On the other hand, $Ri = 1$ is attributed as pure mixed convection heat transfer, referring the equal contribution of both natural and forced convection over the thermal transport phenomena.

Enormous investigations are done both numerically and experimentally considering mixed convective heat transfer inside cavities with various boundary conditions. Perhaps the first attempts to investigate the mixed convection effect in a square cavity evolved from the interplay between buoyancy-driven force and shear-driven force generated from a rotating circular cylinder were taken by Ghaddar and Thiele [7], Fu *et al*. [8], and Kimura *et al*. [9]. Among them, Ghaddar and Thiele [7] analyzed the problem using spectral element method considering the cylinder as a heat source. Fu *et al*. [8] analyzed the problem by setting an isothermal rotating cylinder close to the hot wall and used penalty finite element method [10] to study the effects of mixed convective heat transfer on air as the working fluid. Kimura *et al.* [9] chose water as the working fluid and experimentally investigated the heat transfer performance in the turbulent flow regime. Hsu *et al*. [3] carried out investigation inside a partially heated vented cavity with vertical conductive baffle considering the effects of height and position of the baffle, position of the heat source and location of the outlet. They inferred that the best heat transfer performance can be achieved by placing the outlet at the lower part of the vertical wall. Manca *et al.* [11] examined the effects of hot wall location over the heat transfer characteristics inside a partially opened enclosure utilizing finite element approach, and found that when the hot wall is positioned against the inlet opening, the heat transfer performance becomes the maximum. After that, Manca *et al.* [12] practically examined the heat transfer characteristics inside the same cavity considering only the case where the heated wall was placed near the inlet opening. The outcome revealed that the heat transfer increases by increasing the distance between the two vertical walls. Raji *et al.* [13–15] addressed the case of heated surfaces in a series of studies, in which they considered vented cavities with uniform heat flux to study the heat transfer characteristics numerically. It was found that the highest heat transfer performance can be achieved by injecting air from the bottom side of the cavity. Similar studies were carried out incorporating the effect of thermal radiation over the mixed convective heat transport mechanism [16–19]. Saha *et al.* [20] numerically studied mixed convection phenomenon applying uniform heat flux in a two-dimensional enclosure. They analyzed the effects of different ventilation orientations and concluded that increasing the thermal parameters (*Gr* and *Re*) resulted in an enhanced cooling effectiveness. Omri and Nasrallah [21] numerically compared the mixed convective effects of two different configurations for the inlet and outlet positionings inside an air-filled cavity. They concluded that when the inlet was located near the bottom of the hot wall, and the outlet was positioned near the top of the cold wall, the maximum cooling performance was obtained. Singh and Sharif [22] further extended the research by taking six distinct air inlet and outlet placements into account in a deferentially heated cavity. In their investigation, the highest cooling performance was achieved by placing the inlet and outlet, respectively, close to the lower part of the cold border and the upper part of the heated border.

Heat transfer and flow behavior inside the cavity can be effectively controlled by introducing various passive elements. Therefore, Rahman *et al.* [23] placed a heat generating circular cylinder in a two-dimensional square enclosure and numerically examined the effect of varying Prandtl and Reynolds numbers. The conclusion was that the heat transfer rate can be increased by increasing the thermal parameters (*Re, Pr*), keeping the *Ri* at a particular value. Gupta *et al.*

[24] carried out an analogous analysis applying different boundary conditions, and showed that heat transfer performance increased with increasing the diameter of cylinder and decreased with increasing *Ri*. Rahman *et al.* [25] inserted a heat-generating square block inside a two-dimensional vented cavity. They showed that the size of blocks strongly influenced the flow pattern inside the cavity and for a particular location of the block, the average Nusselt number increased with increasing *Ri*. Gangawane *et al.* [26] also performed similar numerical investigation by placing an adiabatic block of different shapes (triangular, square, circular ) in a semi-circular lid-driven cavity with a heated curved wall with non-Newtonian power-law fluids as working fluid and by varying Richardson number ($0.1 \leq Ri \leq 10$), non-Newtonian power-law index ($0.5 \leq n \leq 1.5$), modified Prandtl number ($1 \leq Pr \leq 100$). They found that increasing the Richardson number and power-law index decrease convective heat transfer whereas, increasing the Prandtl number increases convective heat transfer. They also found triangular blockage reduces heat transfer rate less than other shapes. Similar investigation was done by Chamkha *et al.* [27] in a vented square enclosure, where the square block was isothermally heated. They demonstrated the impact of the aspect ratio of the cylinder and also the change of thermal parameters (*Re* and *Ri*) over the heat transfer performance. Vijayan *et al*. [28] investigated the mixed convection problem numerically inside a lid-driven cavity with a triangular heat source utilizing non-Newtonian power-law fluids as the working fluid. They investigated by changing aspect ratios (AR = length/height = 0.25, 0.5, 0.75), non-Newtonian power-law index ($0.6 \leq n \leq 1.4$), Prandtl number ($1 \leq Pr \leq 100$), and Richardson number ($Ri$ = 0.1, 1, 10). They suggested that for higher aspect ratio and higher power-law index, heat transfer decreases while it increases with the increase in Prandtl number. Also, for a given Richardson number, *Pr* has a more prominent effect on the heat transfer performance compared to the non-Newtonian power-law index. Further, mixed convection heat transfer in a 3D vented cavity with the presence of an isothermal heating block was numerically investigated by Doghmi *et al.* [29]. The parametric simulation was carried out by varying the Reynolds ($50 \leq Re \leq 100$) and Richardson numbers ($0 \leq Ri \leq 10$). The effects of inlet and outlet positioning over the thermal transport phenomenon were analyzed. The investigation revealed that for a particular Richardson number, heat transfer performance increased by increasing *Re* numbers. Furthermore, mixed convection heat transfer with active flow modulation inside a channel via rotating cylinders was explored by Billah *et al*. [30]. The outcome of the investigation was that the configuration type and rotational direction

of cylinder greatly affected the heat transfer in the channel. Selimefendigil and Chamkha [31] examined the heat transfer behavior by placing a heated rotating cylinder in a *CuO*-water nanofluid filled 3D ventilated chamber with corrugated bottom heated surface while an external magnetic field was present. They found that heat transfer rate of the bottom heated surface increases when the cylinder rotates in the counter-clockwise direction. Kimura *et al.* [32] experimentally demonstrated that a rotating blade placed in a square enclosure enhanced the heat transfer performance compared to a rotating cylinder. A similar investigation of mixed convection inside a square cavity having a rotating flat plate was carried out by Lee *et al.* [33] and it was shown that thermal oscillation occurred after Rayleigh number exceeded a threshold value. Sepyani *et al*. [34] employed finite difference method to understand the flow and thermal field characteristics inside an water-$Al_2O_3$ nanofluid filled differentially heated square chamber with an inner rotating blade. They considered the impacts of nanoparticle concentration, length of the rotating blade, Richardson number and Rayleigh number. Islam *et al.* [35] demonstrated the effects of placing periodically distributed rotating flat plates over the enhancement of heat transfer in a long horizontal channel.

From the above literature survey, it can be noticed that no research has been accomplished in a differentially heated ventilated square cavity with rotating flat plate in the light of pure mixed convective heat transfer. The present work aims to extensively study the effects of changing the rotational speed of the blade on the overall heat transfer, inlet pressure and power consumption of the rotating blade. The main attention is given to examine the effects of different velocity ratios, power spectrum analysis and the time averaged Nusselt number.

## *2. Physical Model:*

The physical domain of the present study, as shown in Fig. 1, is a vented square cavity of dimensions *H* and *L(= H)*. The configuration consists of an inlet opening of relative height *h(= 0.1H)* located at the bottom of the left vertical wall, an outlet opening of same relative height placed at the top of the opposite vertical wall and a counterclockwise rotating blade of negligible thickness placed at the center of the cavity. The relative diameter of the blade is *d(= 0.5H)*. Water (*Pr = 7.1*) enters into the cavity with a parabolic velocity profile, where the maximum velocity is denoted by $V_{max}$. The pressure of the outlet is kept constant at *P₀*. The velocity of the blade at blade tip is $v_{blade}(= \omega d/2)$, where $\omega$ is the angular velocity of the rotating blade. The velocity ratio (*r*) is defined as the ratio of the instantaneous blade tip velocity, $(v_{blade})$ to the maximum velocity at the inlet, $(V_{max})$. The top and bottom walls of the enclosure are assumed at a constant lower temperature ($T_c$) and higher temperature ($T_h$) respectively, while the rest of the boundaries are considered as insulated.

## 3.1 Mathematical Model

To avoid complexity, the simulation is performed in the light of taking several assumptions into account. For instance, the fluid motion inside the chamber is taken to be laminar, incompressible, Newtonian, transient, and two-dimensional. Boussinesq approximation is employed for coupling between flow and thermal fields, and to address the changes in density of water with temperature, while other thermophysical properties of water are thought to be independent of temperature variation. All the solid boundaries including the rotor surface are considered to be at no-slip velocity boundary condition. Owing to having an insignificant contribution in outputs, thermal radiation, joule heating, and viscous dissipation effects are omitted from the energy equation.

Taking these considerations into account, within the frame of Cartesian coordinate system defined in Fig.1, Navier Stokes equations along with continuity and energy equations used in the present study are as followed:

$$\frac{\partial u}{\partial x} + \frac{\partial v}{\partial y} = 0 \tag{1}$$

$$\rho \left( \frac{\partial u}{\partial t} + u \frac{\partial u}{\partial x} + v \frac{\partial u}{\partial y} \right) = -\frac{\partial p}{\partial x} + \mu \left( \frac{\partial^2 u}{\partial x^2} + \frac{\partial^2 u}{\partial y^2} \right) \tag{2}$$

$$\rho \left( \frac{\partial v}{\partial t} + u \frac{\partial v}{\partial x} + v \frac{\partial v}{\partial y} \right) = -\frac{\partial p}{\partial y} + \mu \left( \frac{\partial^2 v}{\partial x^2} + \frac{\partial^2 v}{\partial y^2} \right) + \rho g \beta (T - T_c) \tag{3}$$

$$\rho C_p \left( \frac{\partial T}{\partial t} + u \frac{\partial T}{\partial x} + v \frac{\partial T}{\partial y} \right) = k \left( \frac{\partial^2 T}{\partial x^2} + \frac{\partial^2 T}{\partial y^2} \right) \tag{4}$$

To solve these equations numerically utilizing finite element method, equations (1) - (4) are normalized and a non-dimensional form is achieved as followed:

$$\frac{\partial U}{\partial X} + \frac{\partial V}{\partial Y} = 0 \qquad (5)$$

$$\frac{\partial U}{\partial \tau} + U\frac{\partial U}{\partial X} + V\frac{\partial U}{\partial Y} = -\frac{\partial P}{\partial X} + \frac{1}{Re}\left(\frac{\partial^2 U}{\partial X^2} + \frac{\partial^2 U}{\partial Y^2}\right) \qquad (6)$$

$$\frac{\partial V}{\partial \tau} + U\frac{\partial V}{\partial X} + V\frac{\partial V}{\partial Y} = -\frac{\partial P}{\partial Y} + \frac{1}{Re}\left(\frac{\partial^2 V}{\partial X^2} + \frac{\partial^2 V}{\partial Y^2}\right) + Ri\theta \qquad (7)$$

$$\frac{\partial T}{\partial \tau} + U\frac{\partial T}{\partial X} + V\frac{\partial T}{\partial Y} = \frac{1}{RePr}\left(\frac{\partial^2 \theta}{\partial X^2} + \frac{\partial^2 \theta}{\partial Y^2}\right) \qquad (8)$$

The succeeding dimensionless parameters are introduced for obtaining the normalized form of the above governing equations (5) – (8):

$$X = \frac{x}{H}, Y = \frac{y}{H}, U = \frac{u}{v_{max}}, V = \frac{v}{v_{max}}, P = \frac{p - p_o}{\rho v_{max}^2}, \theta = \frac{T - T_c}{T_h - T_c}, \tau = \frac{t v_{max}}{H}$$

The non-dimensional numbers corresponding to the present problem are Reynolds number (*Re*), Grashof number (*Gr*), Richardson number (*Ri*), and Prandtl number (*Pr*), which are mathematically represented in the following forms:

$$Re = \frac{\rho v_{max} H}{\mu}, Gr = \frac{g\beta(T - T_c)H^3 \rho^2}{\mu^2}, Ri = \frac{Gr}{Re^2}, Pr = \frac{\mu}{\rho\alpha}$$

Table 1 describes the thermal and velocity boundary conditions in dimensionless form corresponding to the current numerical model.

Table 1: Non-dimensional boundary conditions.

| Boundary | Thermal field | Velocity field |
|---|---|---|
| Top wall | $\theta = 0$ | $U = V = 0$ |
| Bottom wall | $\theta = 1$ | $U = V = 0$ |
| Vertical walls | $\partial\theta/\partial X = 0$ | $U = V = 0$ |

The performance of mixed convection heat transfer inside the cavity is evaluated from the spatially averaged ($Nu_{avg}$) and the time averaged ($Nu_\tau$) Nusselt number calculated along the hot wall. The spatially averaged and time averaged Nusselt numbers are defined as below:

$$Nu = -\left(\frac{\partial \theta}{\partial Y}\right)_{Y=0} \tag{9}$$

$$Nu_{avg} = \int_0^1 Nu\, dX \tag{10}$$

$$Nu_\tau = F_t \int_0^{1/F_t} Nu_{avg}\, d\tau \tag{11}$$

Here, $F_t$ is the frequency of the spatially averaged Nusselt number obtained by Fast Fourier Transform (FFT) analysis. The time averaged Nusselt number is evaluated for a single cycle of Nusselt number after the system reaches steady periodic spatially averaged Nusselt number. The periodicity of $Nu_{avg}$ will be discussed in the next sections. The spatially averaged and time averaged power required by the rotating blade against the fluid motion are defined as:

$$\dot{W}_{avg} = -\int P v_n\, ds \tag{12}$$

$$\dot{W}_\tau = F_w \int_0^{1/F_w} \dot{W}_{avg}\, d\tau \tag{13}$$

Here, $F_w$ is the frequency of the spatially averaged power obtained by FFT analysis and $v_n$ is the velocity normal the rotating blade. The non-dimensional spatially averaged pressure ($P_{avg}$) of the inlet at a particular time ($\tau$), time averaged non-dimensional inlet pressure ($P_\tau$) and the developed pressure difference ($\Delta P$) between the outlet ($P_0$) and the inlet ($P_\tau$) are calculated as:

$$P_{avg} = \frac{1}{h}\int_0^h P_{X=0}\, dY \tag{14}$$

$$P_\tau = F_p \int_0^{1/F_p} P\, d\tau \tag{15}$$

$$\Delta P = P_o - P_\tau \tag{16}$$

Symbols presented in the above equations (1)-(16) are introduced with proper details in the nomenclature section.

*3.2 Numerical Procedure:*

The governing equations (5)-(8) and the corresponding boundary conditions (Table-1) are modeled and solved using COMSOL® Multiphysics commercial package [36], which basically performs numerical analysis utilizing finite element method. The moving mesh problem is formulated in COMSOL by using Arbitrary Lagrangian Eulerian (ALE) finite element formulation. The application of this method for solving the fluid-solid system for moving bodies was documented by Glowinsky *et al.* [37] and Hu *et al.* [38]. The governing equations are resolved with a nonlinear parametrial solver which guarantees quicker convergence and reliability. In the present study, non-uniform quadrilateral mesh elements are implemented. The elements are refined near the computational domain boundaries as well as near the surface of the blade as rapid changes of dependent variables are more prominent near these regions. Figure 2 represents the meshed geometry corresponding to a blade of length $d = 0.5$ at the center of the cavity.

*3.3 Grid Sensitivity Test:*

To find an ideal grid distribution providing results with reasonable accuracy leveraging optimum computational time, several numerical simulations have been done with varying the element size and the corresponding outputs are listed in Table 2, in the form of time-averaged Nusselt number.

This test is performed for Reynolds number ($Re$) = 100, Richardson number ($Ri$) = 1, Prandtl number ($Pr$) = 7.1, and velocity ratio ($r$) = 2. It is evident from Table 2 that increasing element number higher than 5048 does not cause any significant impact over the time-averaged Nusselt number and hence, mesh element 5048 is used for carrying out the entire numerical analysis of the present study.

Table 2: Comparison of time averaged Nusselt numbers along the bottom heated wall for various grid dimensions.

| Nodes (Elements) | $Nu_\tau$ |
|---|---|
| 5338(2422) | 2.6298 |
| 6568 (3016) | 2.5725 |
| 8724 (4056) | 2.5159 |
| **10784 (5048)** | **2.4979** |
| 14144 (6672) | 2.4836 |
| 17240 (8182) | 2.4718 |

*3.4 Model Validation:*

The present model has been verified against the published result of Chamkha *et al.* [27]. The comparison is done based on the calculated average Nusselt number along the heated boundary with the variation of Richardson number, and Fig. 3 depicts the comparison results. In this case, the inlet is always positioned at the middle of the left cold vertical border, whereas, according to the outlet location in the right adiabatic vertical border, three different configurations are considered, which are CT configuration (inlet in center, outlet in top), CC configuration (inlet in center, outlet in center), and CB configuration (inlet in center, outlet in bottom). In Fig. 3, Solid lines represent values obtained from present simulation procedure and dashed line represents values evaluated by Chamkha *et al.* [27]. From this comparison, it can be concluded that the present model and the numerical procedure provide the computational results with reasonable accuracy.

Moreover, the present numerical procedure of moving mesh has been compared with the existing results of Lee *et al.* [33]. The variation of Nusselt number with dimensionless time for $d = 0.6$, and $Re = 430$ for four representative Rayleigh numbers ($Ra$) have been compared, as presented in Fig. 4. The consistency of the present simulation outputs with the prior studies provides enough confidence to ensure the precision of the current simulation process.

*4. Results and Discussion*

Results found in the present numerical simulation for laminar pure mixed convection ($Ri = 1$) inside a ventilated square enclosure filled with water ($Pr = 7.1$) having a rotating blade with negligible thickness at the center has been discussed here. In the present system, computation started from a random initial state. When a steady state is achieved, non-dimensional time is set to $\tau = 0$ and then the computation has been carried out for another 300 cycles ($0 \leq \tau \leq 300$). The effect of varying velocity ratio ($r$) of the rotating blade on mixed convective flow and thermal transport characteristics has been extensively analyzed. We kept the Reynolds number ($Re$) at a fixed value of 100 and varied the velocity ratio ($r$) from 1 to 10. Streamlines and isothermal contours plots are incorporated to understand the flow pattern and thermal field distribution inside the domain.

The time-averaged Nusselt number, the inlet-outlet pressure difference, the power required for the rotating blade has been plotted against the velocity ratio to demonstrate their impacts over the heat transfer performance. The effect of different velocity ratios over the heat transfer performance of the heated bottom wall is described by evaluating the time-averaged Nusselt number ($Nu_\tau$). The frequency of the spatially averaged Nusselt number ($Nu_{avg}$) is determined utilizing the Fast Fourier transformation (FFT).

*4.1 Effects on Flow and Thermal fields:*

The effects of velocity ratio (*r*) over the flow and thermal fields inside the cavity are presented via streamlines and isothermal contours in Fig. 5 and Fig. 6, respectively, for $Re = 100$ and $\tau = 220$. This particular non-dimensional time ($\tau = 220$) has been chosen to demonstrate the effects of velocity ratios when the rotating blade is in the horizontal position. It can be seen from Fig. 5 that irrespective of any velocity ratio a primary counterclockwise vortex is generated at the center of the cavity due to the counterclockwise rotation of the blade. Additionally, two clockwise secondary vortices at the left top and bottom right corner of the cavity are also present due to the shear effect resulted from the large velocity difference between the fluid around the rotating blade tip and the almost stationary fluid adjacent to these corners of the cavity. When *r* = 1, no such eddies are seen near the inlet and outlet port as the fluids near these two locations are in motion. However, as *r* increases, due to the increased velocity difference between fluids near the blade tip and the fluids near these two openings, secondary eddies also appear in those locations. Moreover, the size of the secondary vortices increases with the increasing velocity ratio because of the increased shearing effect between the fluids. The strength of the counterclockwise vortex at the cavity core increases with the increment in velocity ratios and it gradually expands to the sidewalls of the cavity and thus, reduces the hydrodynamic boundary layer thickness.

From Fig. 6, it is evident that the thermal field is also greatly affected by the increment in the speed of the rotating blade. For low velocity ratios, the isothermal contours are spread across the computational domain. As velocity ratio increases, the hotter isotherms get pushed near the center of the bottom wall and the colder isotherms get clustered near the top-right corner of the cavity due to the increased strength of the counterclockwise circulation. So, the thickness of the thermal boundary layer decreases, and the temperature gradients increase at these locations with increased blade velocity, which results in an enhanced heat transfer. Also, the congested parallel isotherms near the bottom and the top walls represent the prominence of conduction heat transfer near these walls. As, no heat transfer is taking place between the fluid and the vertical adiabatic walls, isothermal lines are perpendicular to these walls.

*4.2. Effects on Nusselt numbers and developed pressure:*

Variation of time-averaged Nusselt number ($Nu_\tau$) as a function velocity ratio is shown in Fig. 7. Note that, all the values of the Nusselt number presented in this paper are calculated along the bottom heated wall. It is evident that increasing the velocity ratio (*r*) has a profound effect on enhancing heat transfer and as such $Nu_\tau$ increases linearly with the increase of *r*. In fact, increasing the velocity ratio from 1 to 10 causes an increment in the $Nu_\tau$ of about 127.3%. This can be attributed to the fact that increasing the relative velocity of the blade strengthens the forced convection effect and promote better mixing between the hot and cold fluid and increase the temperature gradient near the hot wall (Fig. 6). In some previous studies, rotating blades and cylinders [39], in certain cases caused a blockage effect, which reduced the heat transfer. However, in the present model, no such blockage effect has been observed.

Figure. 8. shows to the effect of the velocity ratio over the developed pressure difference between the outlet and the inlet. Increasing velocity ration from 1 to 10, causes significant increment in the developed pressure. As the blade is rotating counterclockwise it induces a suction effect at the inlet and pushes the fluid toward the outlet. So, a low-pressure zone is created in front of the inlet and a high-pressure zone behind the outlet. As the relative velocity of the blade increases the suction and pushing effects increases which eventually results in a large pressure difference. Hence it can be concluded that, by changing the velocity ratio (*r*), we can modulate the developed pressure difference and thus the flow characteristics. Figure 7 and 8 also show that, the developed pressure and time averaged Nusselt number are roughly proportional to each other, which affirms that developed pressure in a ventilated cavity with rotating blade has significant effect on the heat transfer characteristics.

*4.3. Effects on thermal frequency, blade frequency and inlet pressure frequency*

In Fig. 9, it can be seen that the spatially averaged Nusselt number ($Nu_{avg}$) representing the heat transfer characteristics of the bottom wall, shows oscillating behavior for different values of velocity ratio (*r*). It is quite obvious as the continuous rotation of the blade causes a periodic contraction and expansion effect in the thermal boundary layer near the bottom wall. When the boundary layer contracts the temperature gradient increases which results in a higher heat transfer rate. On the contrary, when the thermal boundary layer expands the temperature gradient between the hot wall and the adjacent fluid layer decreases and the heat transfer, hence, the Nusselt number decreases. As the peripheral speed of the blade increases, the $Nu_{avg}$ increases due to the increased forced convection effect. Also, it can be noted that the amplitude of oscillation continuously decreases with higher speed ratios. This can be described by the fact that when the blade rotates at a lower speed, the thermal boundary layer adjacent to the hot wall gets enough time for expansion and contraction. So, the variation in the $Nu_{avg}$ is noticeable. On the other hand, when the blade rotates at a higher speed, the boundary layer does not get enough time for expansion and contraction. This type of oscillation of Nusselt number due to blade rotation has been observed in previous studies [33,35].

Figure 10(a) and 11(a) represents a magnified view of the variation of $Nu_{avg}$ with time for different velocity ratios of the rotating blade. From these figures, it is clear that the $Nu_{avg}$ does not follow any perfect sinusoidal patterns, rather they resemble various complex shapes. Also, a periodic pattern in the fluctuation of $Nu_{avg}$ is observed up to $r = 4$, beyond that the variation of $Nu_{avg}$ does not follow any specific trend. To find out the frequencies of thermal fluctuation from these plots, power spectrum analysis is done as shown in Fig 10(b) and 11(b) with the aid of Fast Fourier Transformation (FFT), and some interesting results are observed. The x-axis of the power spectrum plot represents all the potential frequencies of the oscillating curve of $Nu_{avg}$, and the y-axis represents the intensity of $Nu_{avg}$ at that particular frequency. It is seen that when $r < 4$, the oscillating waveform can be described by a single frequency along with its integer multiples. It indicates that the oscillation has a particular time period. However, for $r = 4$, the oscillation does not show any single frequency value. This represents that at a higher rotational speed, the fluctuating heat transfer rate does not show any specific time period or frequency.

Another important observation is that when $r < 4$, the thermal fluctuation frequency is the same as the blade rotation frequency. The rotational frequency of the blade $f_b$ can be calculated from the formula, $v_{blade} = \pi d f_b$ When the velocity ratio, $r = 1$, the non-dimensional frequency ($F_b$) of the blade rotation is 0.6366. Also, from the power spectrum analysis for $r = 1$, it is seen that the frequency of the thermal fluctuations is 0.6348 or its multiple. Similarly, when $r = 2$, the frequency of blade rotation is 1.2732, and the frequency of thermal fluctuation from power spectrum is 1.2744 and its multiples. These represent up to $r = 4$, the flow field remains at thermally stable condition. At $r = 4$, the frequency of the blade is 2.5464, whereas the power spectrum plot gives two independent thermal frequencies that are 2.5488 and 4.9072. This indicates a transition from the stabilized thermal frequency region. Beyond $r = 4$, no specific pattern in the thermal frequency is observed that may be due to complex interaction between blade and fluid flow.

The temporal variation of inlet pressure with varying relative velocity of the blade is depicted in Fig. 12 and Fig. 13. It is clear that the fluctuating inlet pressure profile and the corresponding power spectrum plot represents exactly the same behavior described in Fig. 10 and Fig. 11. Due to the blade rotation a continuous periodic high pressure and low pressure region are created in front of the inlet opening, as described earlier.

Table 3. A comparison of blade frequencies with thermal frequencies and pressure frequencies for different velocity ratios.

| Velocity ratios ($r$) | Blade frequency ($F_b$) | Frequencies of $Nu_{avg}$ from power spectrum plot ($F_t$) | Frequencies of developed pressure from power spectrum plot ($F_p$) |
|---|---|---|---|
| 1 | 0.6366 | 0.6348, 1.2744 | 0.6348, 1.2744 |
| 2 | 1.2732 | 1.2744, 2.5488 | 1.2744, 2.5468 |
| 4 | 2.5464 | 2.5488, 4.9072 | 2.5468, 4.9072 |
| 5 | 3.1830 | 3.1836, 3.6328 | 3.1836, 3.6528 |
| 8 | 5.0930 | 0.1855, 2.1094, 4.9072 | 0.18, 4.9072 |
| 10 | 6.3661 | 0.9033, 2.7344, 3.6328 | 2.7344, 3.6328 |

From Table 3, the changing trend of the thermal frequencies and developed pressure with velocity ratio ($r$) can be understood more profoundly. It is evident that when $r < 4$, both the $Nu_{avg}$ and inlet pressure profile shows the same frequency profile as the blade frequencies. Also, the frequencies are integer multiple to each other. At $r = 4$, two different independent frequencies can be seen. Among them one frequency is the same as the blade frequency. So, it can be said that $r = 4$, is the blade rotational speed from which the stable thermal and flow behavior transits to the unstable region. Beyond $r = 4$, neither the frequencies remain the same as the blade rotational frequency, nor they are an integer multiple to each other. So, a specific frequency within this time domain cannot be obtained for these high rotating speeds. This occurs due to instability of the thermal and hydrodynamic boundary layers that arises from the increased turbulence effect at higher blade rotation speed. When the blade rotates at a high velocity, the sudden expansion and contraction of the thermal and hydrodynamic boundary layers, and hence, the sudden change in the heat and mass transfer rates give rise to these instabilities. Note that the mean value of the time-averaged Nusselt number increases with increasing $r$. This observation is similar to that reported by Chamkha *et al.* [27], although the boundary conditions and configuration of the cavity is different.

## *4.4 Effects on time averaged power*

Increasing the velocity ratio enhances the heat transfer characteristics as shown in Fig. 7. However, with the increasing velocity ratio, the required power to rotate the blade also increases. Figure 14 shows the variation of time-averaged power ($\dot{W}_{avg}$) required to rotate the blade inside the cavity as a function of $r$. It is found that up to $r = 4$, the $\dot{W}_{avg}$ shows a very small dependence on the velocity ratio ($r$). After $r = 4$, the $\dot{W}_{avg}$ increases exponentially with increasing $r$. To further illustrate the relationship between $\dot{W}_{avg}$ and $r$, a fitted curve has been added in Fig. 14. It is clear from the curve that, $\dot{W}_{avg}$ is proportional to the cube of velocity ratio, $r^3$. This increase in $\dot{W}_{avg}$ with $r$, as described earlier, can be attributed to the increased turbulence and instability in the flow field. One of the key observations from the figure is that although the heat transfer increases with $r$, the power requirement after a certain value will be too high to be beneficial for any practical application.

## *5. Conclusion:*

The present numerical model details the combined buoyancy and shear effect over the unsteady laminar heat transfer phenomena inside a ventilated square cavity having a thin rotating blade positioned at the middle acting as active flow modulator. The analysis is performed numerically rendering the use of Arbitrary Lagrangian-Eulerian (ALE) finite element formulation. To understand the consequences of velocity ratio of the rotating blade on the fluid motion and heat transfer characteristics inside the cavity, an in-depth analysis has been carried out by examining the streamlines, isotherms, spatially averaged Nusselt number, time-averaged Nusselt number, the power required by the rotating blade, and average inlet pressure. The major outcomes of the present investigation can be summarized as follows:

- Time-averaged Nusselt number and the pressure difference between inlet and outlet opening increases almost linearly with increasing velocity ratio.
- The spatially averaged Nusselt number shows an oscillating behavior with time due to the periodic contraction and expansion of the thermal boundary layer.
- The amplitude of oscillation continuously decreases with increasing speed ratio.
- Up to $r = 4$, a periodic pattern in the fluctuation of $Nu_{avg}$ and inlet pressure is observed. The oscillating pattern has a single frequency value which is identical to the frequency of the blade rotation.
- Beyond $r = 4$, due to the increased instability in the thermal and flow fields, the oscillation does not show any single thermal frequency value, and also the thermal frequencies are the same to the frequency of blade rotation.
- The time-averaged power ($\dot{W}_{avg}$) required to rotate the blade is proportional to the cube of velocity ratio ($r^3$).

We envisage that this work would assist researchers as well as engineers in the efficient design and thermal management of such enclosures having innumerable applications in various cooling and mixing technologies.

*6. Acknowledgement:*

The authors would like to acknowledge Multiscale Mechanical Modeling and Research Networks (MMMRN) for their technical support to carry out the research. The authors would also like to acknowledge Department of Mechanical Engineering, Bangladesh University of Engineering & Technology (BUET) for providing computational facilities required for this research.

| **Nomenclature** | | $U$ | dimensionless horizontal velocity = $u / v_{max}$ |
|---|---|---|---|
| $C_p$ | Specific heat capacity (J/(kg.K)) | $v$ | vertical velocity component (*m/s*) |
| $D$ | diameter of the blade (*m*) | | |
| $F_b$ | Non-dimensional blade frequency | $V$ | dimensionless vertical velocity = $v / v_{max}$ |
| $F_t$ | frequency of spatially averaged Nusselt number | $v_{max}$ | maximum velocity at inlet (*m/s*) |
| $F_w$ | frequency of spatially averaged power | $\dot{W}_{avg}$ | spatially averaged power |
| $F_p$ | frequency of spatially averaged pressure at inlet | $\dot{W}_\tau$ | time averaged power |
| $g$ | gravitational acceleration (*m/s²*) | $x,y$ | Cartesian co-ordinates (*m*) |
| $Gr$ | Grashof number | $X,Y$ | dimensionless Cartesian co-ordinates |
| $h$ | height of inlet (*m*) | | |
| $H$ | height of cavity (*m*) | ***Greek Symbols*** | |
| $k$ | thermal conductivity (W/(m.K)) | | |
| $L$ | length of cavity (*m*) | $\alpha$ | thermal diffusivity of fluid (*m²/s*) |
| $Nu$ | Nusselt number | $\beta$ | thermal expansion coefficient (*1/K*) |
| $Nu_{avg}$ | spatially averaged Nusselt number | $\omega$ | angular velocity of the rotating blade (rad/s) |
| $Nu_\tau$ | Time averaged Nusselt number | $\mu$ | the viscosity of fluid (*m²/s*) |
| $p$ | fluid Pressure (*Pa*) | $v_{blade}$ | velocity of the blade at blade tip = $\omega d/2$ |
| $P$ | dimensionless fluid pressure $(p-p_o) / \rho v_{max}^2$ | $\theta$ | dimensionless temperature |
| $P_{avg}$ | dimensionless spatially averaged pressure | $\rho$ | working fluid (*kg/m³*) |
| $P_o$ | fluid pressure (*Pa*) at outlet | $\tau$ | dimensionless time |
| $Pr$ | Prandtl Number = $\mu / \rho\alpha$ | | |
| $P_\tau$ | dimensionless time averaged pressure | ***Subscripts*** | |
| $Re$ | Reynolds Number = $\rho v_{max} H / \mu$ | $c$ | Cold |
| $Ri$ | Richardson Number | $h$ | Hot |
| $r$ | velocity ratio = $v_{blade} / v_{max}$ | | |
| | | ***Abbreviates*** | |
| $t$ | time (*s*) | | |
| $T$ | temperature (*K*) | ALE | Arbitrary Lagrangian-Eulerian |
| $u$ | horizontal velocity component (*m/s*) | FFT | Fast Fourier Transformation |

| | List of Figure Captions |
|---|---|
| **Figure 1** | Schematic diagram of the computational domain along with boundary conditions |
| **Figure 2** | Mesh Distribution for the computational domain |
| **Figure 3** | Comparison of the present model with the results of Chamkha *et al.* [27] in terms of average Nusselt number along the hot wall as a function of Ri for different outlet configurations |
| **Figure 4** | Comparison of the variation of *Nu* with time between Lee *et al.* [33] and present model, at $Re = 430$ and $d = 0.6$ |
| **Figure 5** | Comparison of flow fields for Water ($Pr = 7.1$) at $Re = 100$ and $Ri = 1$ for different velocity ratio at $\tau = 220$ |
| **Figure 6** | Comparison of Isothermal contours of Water ($Pr = 7.1$) at $Re = 100$ and $Ri = 1$ for different velocity ratio at $\tau = 220$ |
| **Figure 7** | Variation of time average Nusselt Number with velocity ratio for $Re = 100$, $Ri = 1$, $Pr = 7.1$ |
| **Figure 8** | Variation of Inlet and outlet pressure difference with velocity ratio for $Re = 100$, $Ri = 1$, $Pr = 7.1$ |
| **Figure 9** | Variation of spatially averaged Nusselt number with non-dimensional time for different velocity ratio at $Re = 100$, $Ri = 1$, $Pr = 7.1$ |
| **Figure 10** | (a) Temporal variation of spatially averaged Nusselt number, (b) One sided fast Fourier transform power spectrum for different velocity ratio ($r = 1, 2, 4$) at $Re = 100$, $Ri = 1$, $Pr = 7.1$ |
| **Figure 11** | (a) Temporal variation of spatially averaged Nusselt number, (b) One sided fast Fourier transform power spectrum for different velocity ratio ($r = 5, 8, 10$) at $Re = 100$, $Ri = 1$, $Pr = 7.1$ |
| **Figure 12** | (a) Temporal variation of inlet pressure, (b) One sided fast Fourier transform power spectrum of inlet pressure for different velocity ratio ($r = 1, 2, 4$) at $Re = 100$, $Ri = 1$, $Pr = 7.1$ |
| **Figure 13** | (a) Temporal variation of inlet pressure, (b) One sided fast Fourier transform power spectrum of inlet pressure for different velocity ratio ($r = 5, 8, 10$) at $Re = 100$, $Ri = 1$, $Pr = 7.1$ |
| **Figure 14** | Variation of time average power ($\dot{W}avg$) with velocity ratio ($r$) at $Re = 100$, $Ri = 1$, $Pr = 7.1$ |

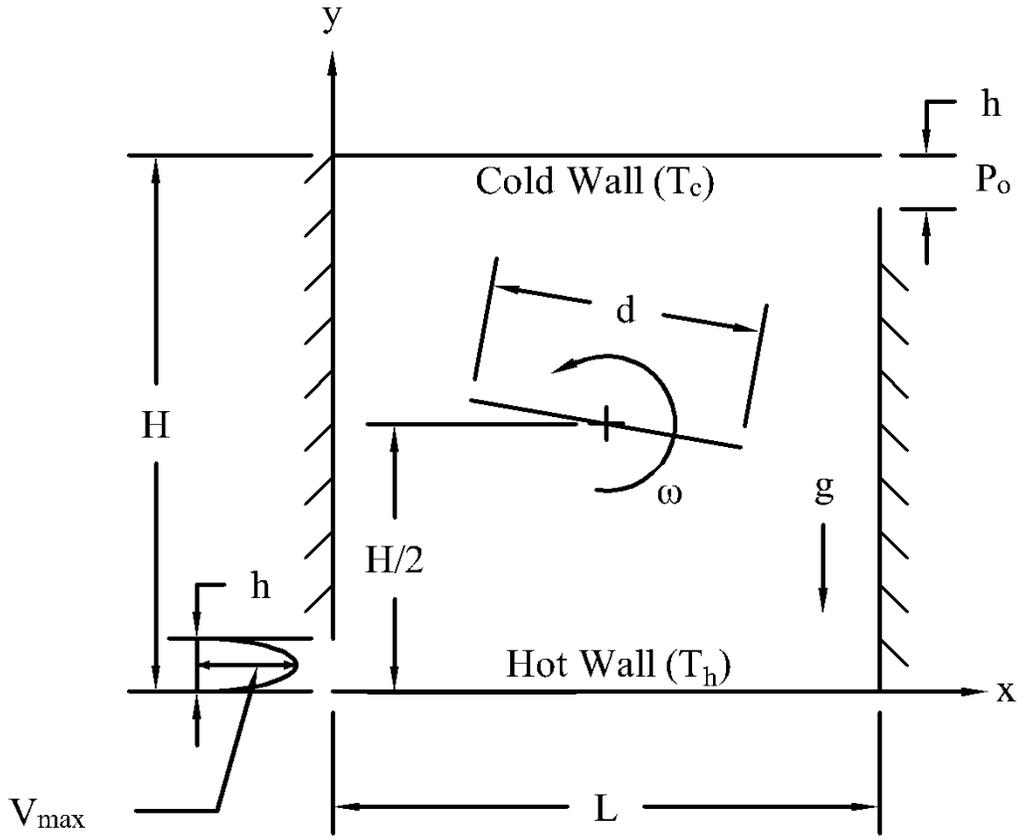

Figure 1. Schematic diagram of the computational domain along with boundary conditions

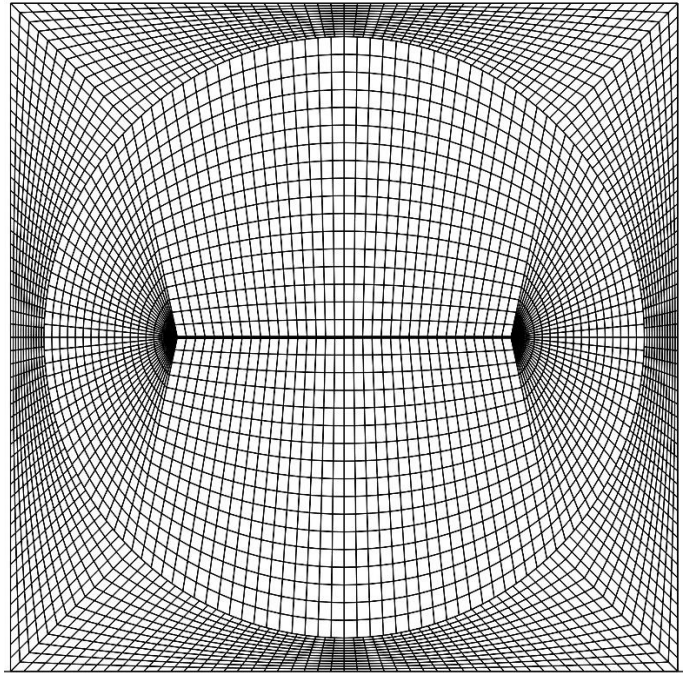

Figure 2. Mesh Distribution for the computational domain

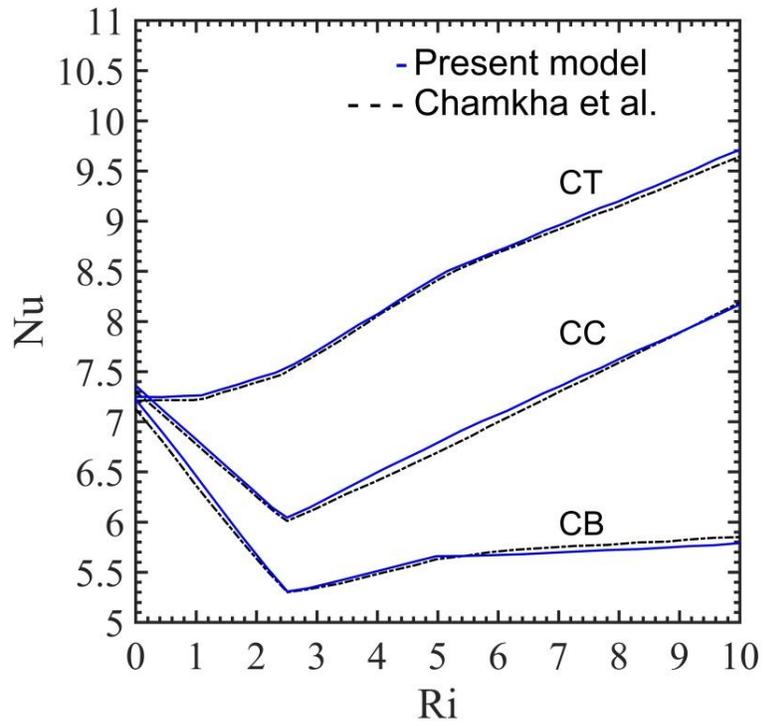

Figure 3. Comparison of the present model with the results of Chamkha *et al.* [27] in terms of average Nusselt number along the hot wall as a function of *Ri* for different outlet configurations

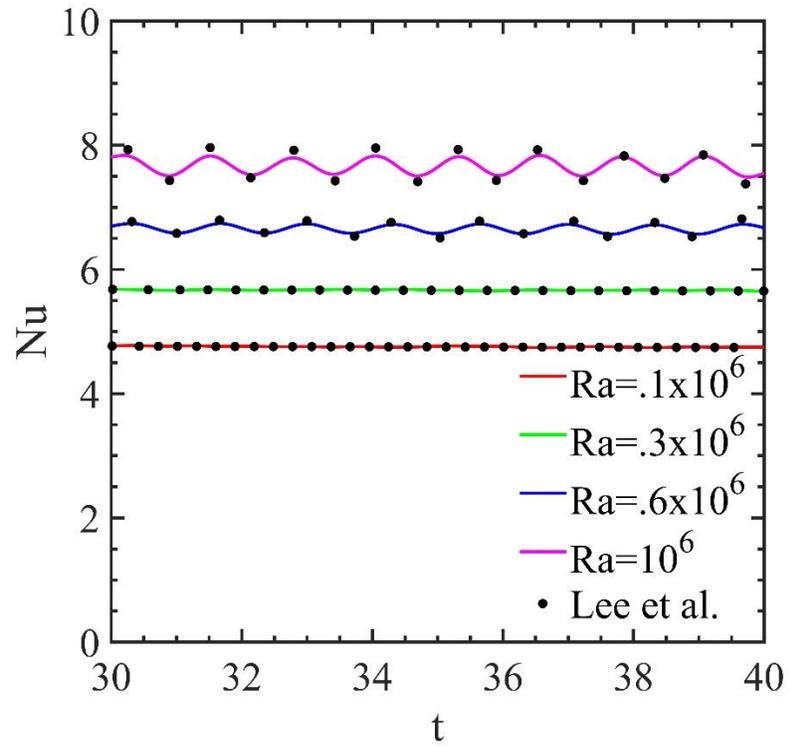

Figure 4: Comparison of the variation of Nu with time between Lee *et al*. [33] and present model, at *Re* = 430 and *d* = 0.6

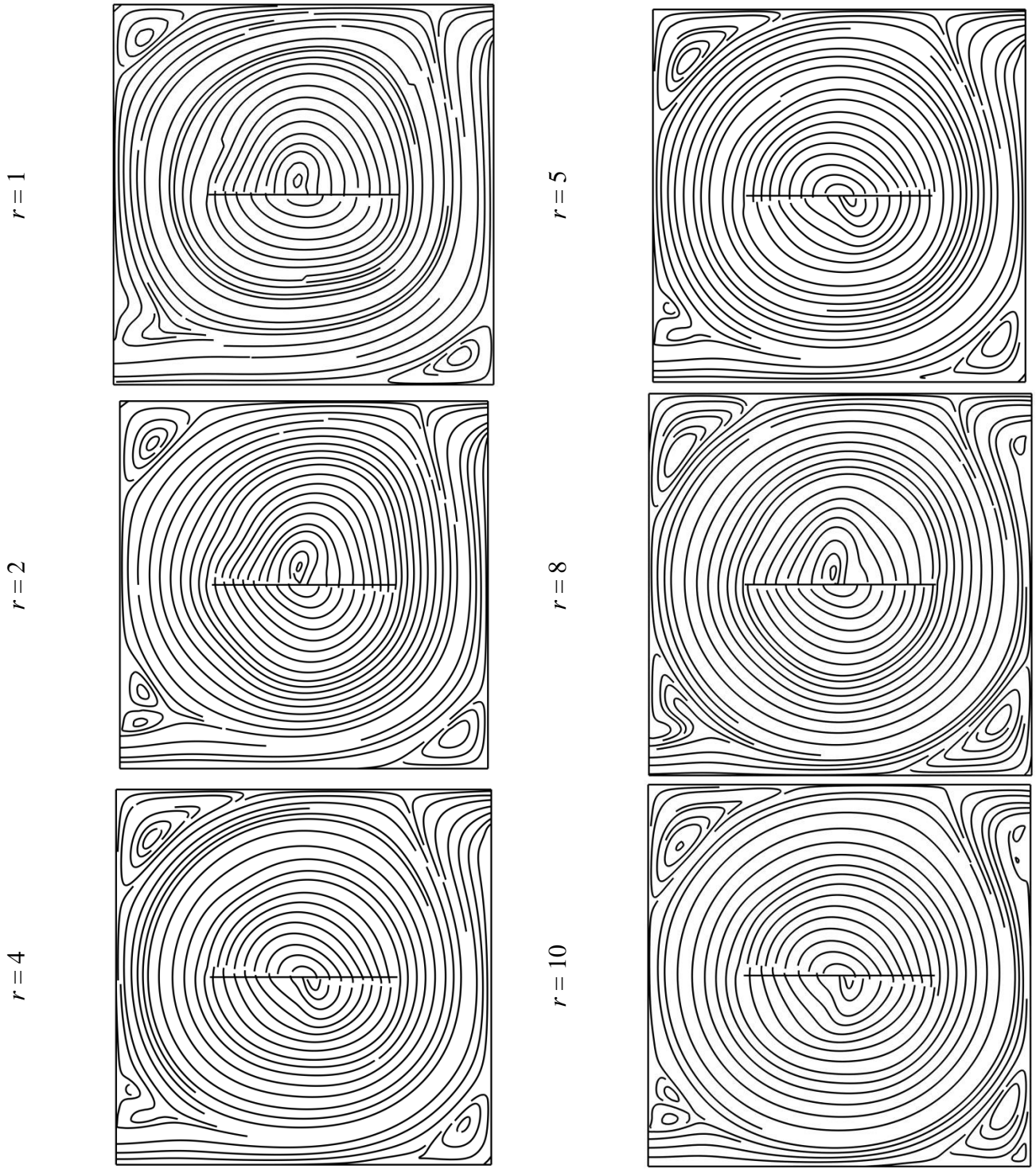

Figure 5. Comparison of flow fields for Water ($Pr = 7.1$) at $Re = 100$ and $Ri = 1$ for different velocity ratio at $\tau = 220$

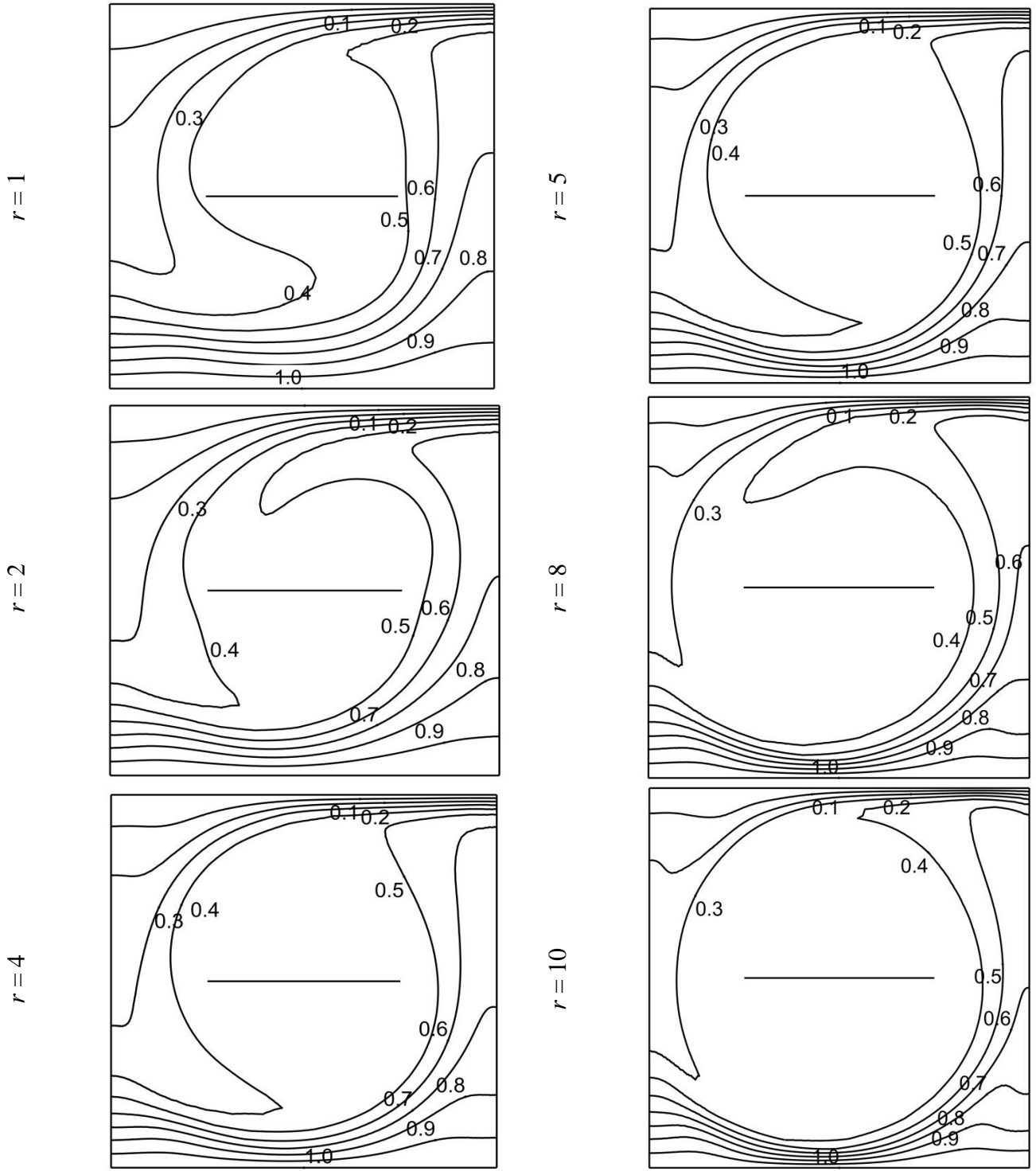

Figure 6. Comparison of Isothermal contours of Water ($Pr = 7.1$) at $Re = 100$ and $Ri = 1$ for different velocity ratio at $\tau = 220$

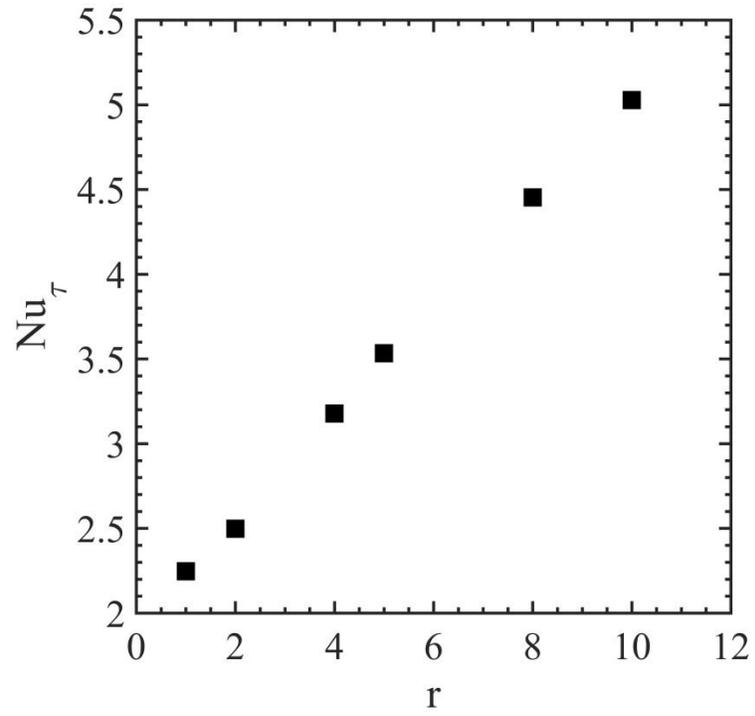

Figure 7. Variation of time average Nusselt Number with velocity ratio for *Re* = 100, *Ri* = 1, *Pr* = 7.1

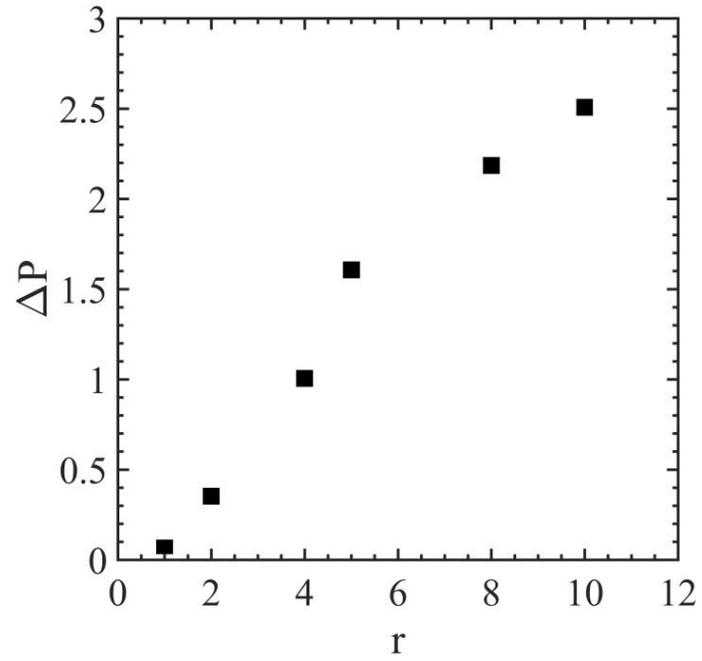

Figure 8. Variation of Inlet and outlet pressure difference with velocity ratio for *Re* = 100, *Ri* = 1, *Pr* = 7.1

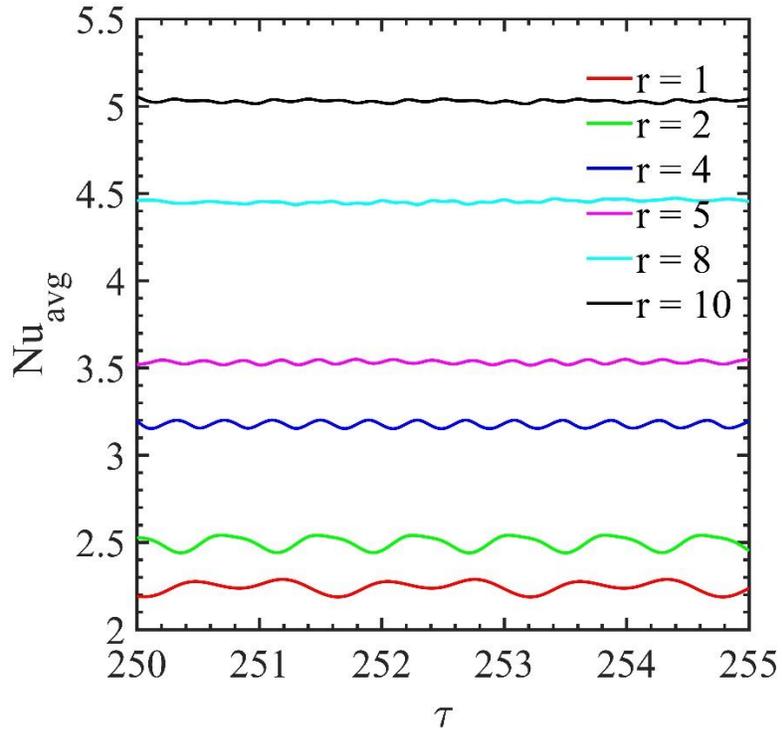

Figure 9: Variation of spatially averaged Nusselt number with non-dimensional time for different velocity ratio at *Re* = 100, *Ri* = 1 , *Pr* = 7.1

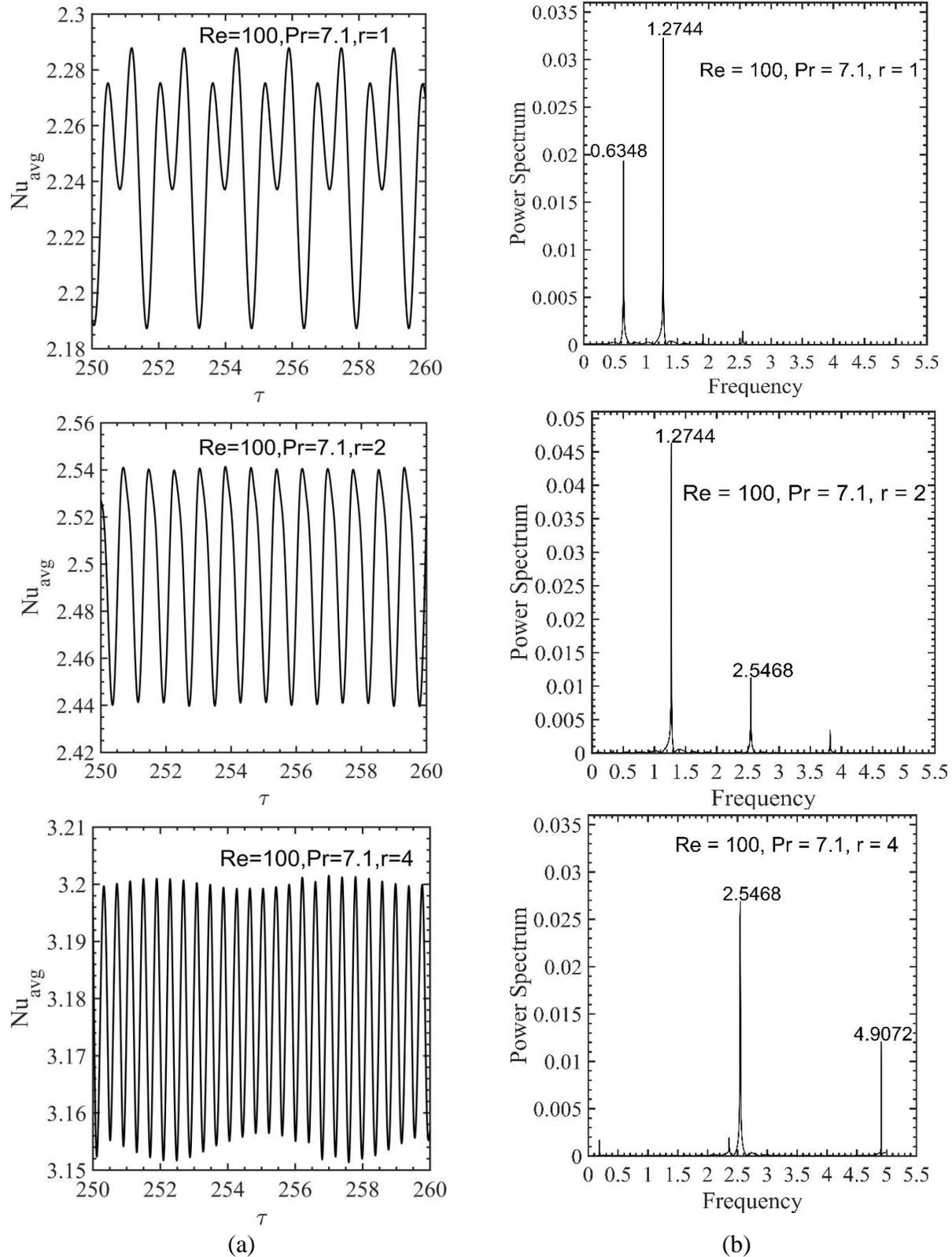

(a) (b)

Figure 10. (a) Temporal variation of spatially averaged Nusselt number, (b) One sided fast Fourier transform power spectrum for different velocity ratio ($r$ = 1, 2, 4) at $Re$ = 100, $Ri$ = 1, $Pr$ = 7.1

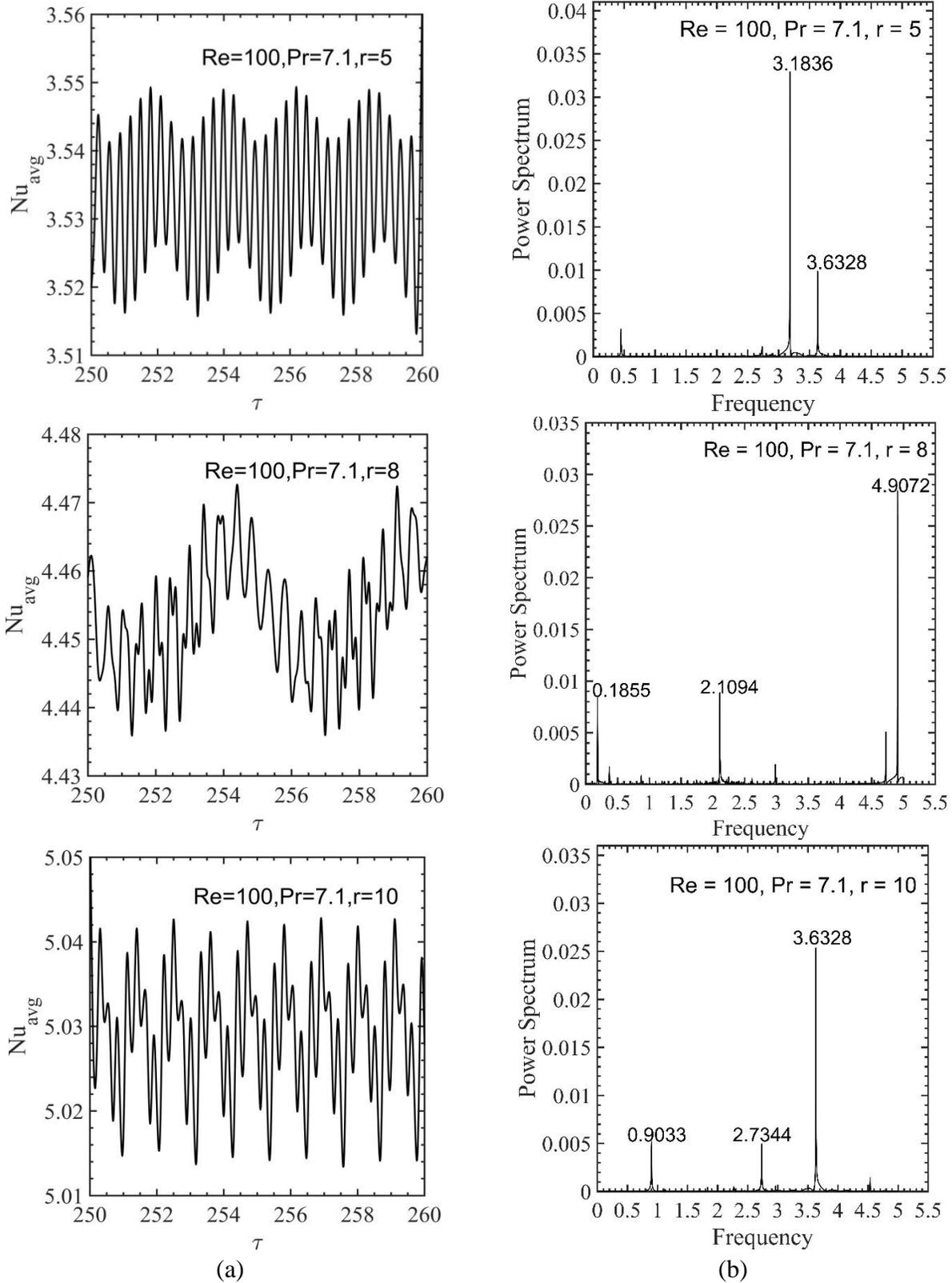

Figure 11. (a) Temporal variation of spatially averaged Nusselt number, (b) One sided fast Fourier transform power spectrum for different velocity ratio ($r = 5, 8, 10$) at $Re = 100$, $Ri = 1$, $Pr = 7.1$

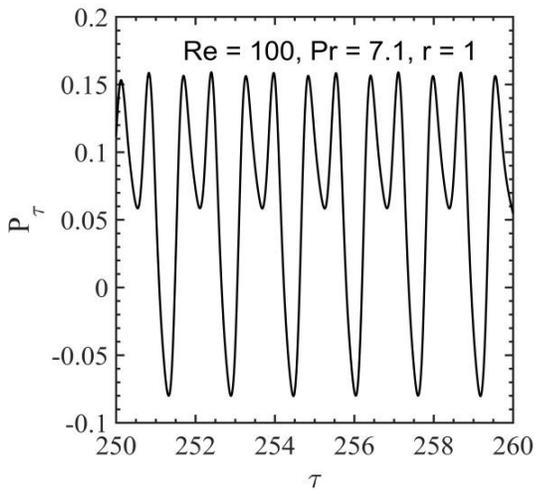
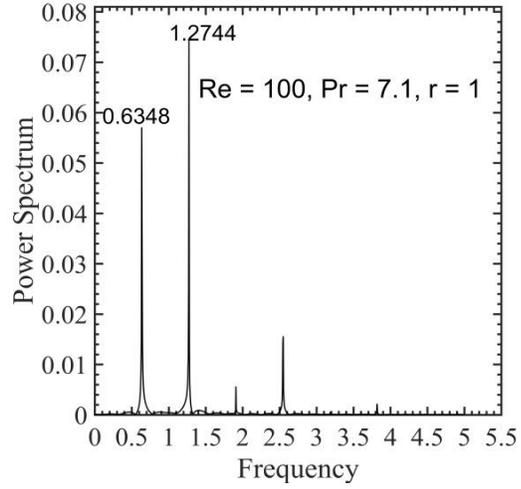
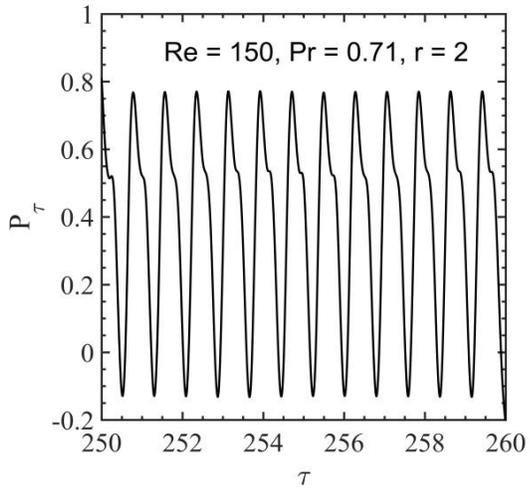
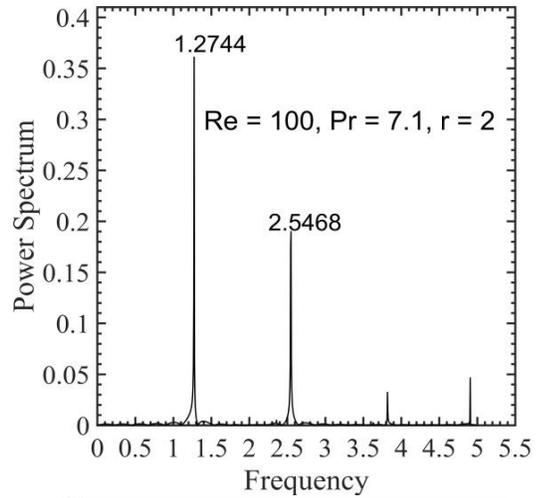
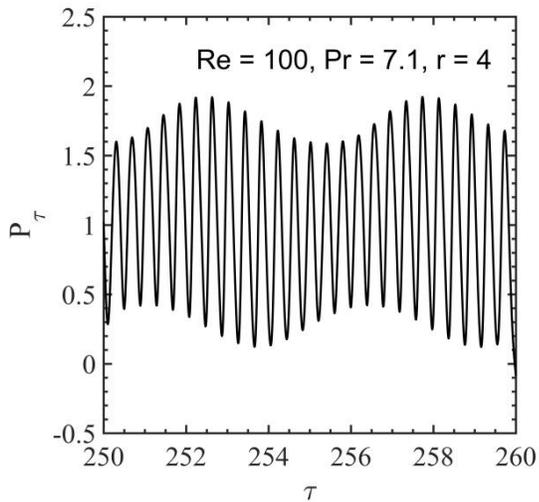
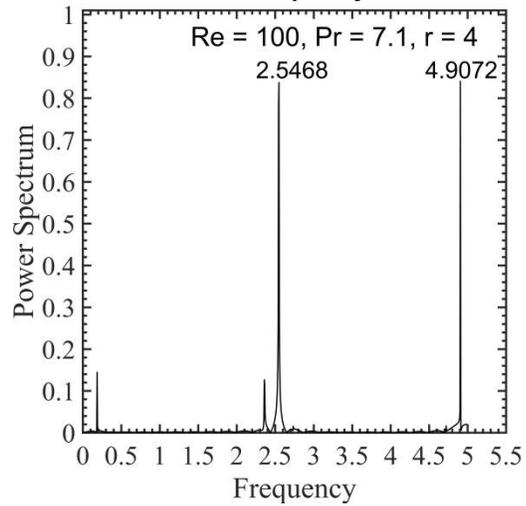

(a)                          (b)

Figure 12. (a) Temporal variation of inlet pressure, (b) One sided fast Fourier transform power spectrum of inlet pressure for different velocity ratio ($r$ = 1, 2, 4) at $Re$ = 100, $Ri$ = 1, $Pr$ = 7.1

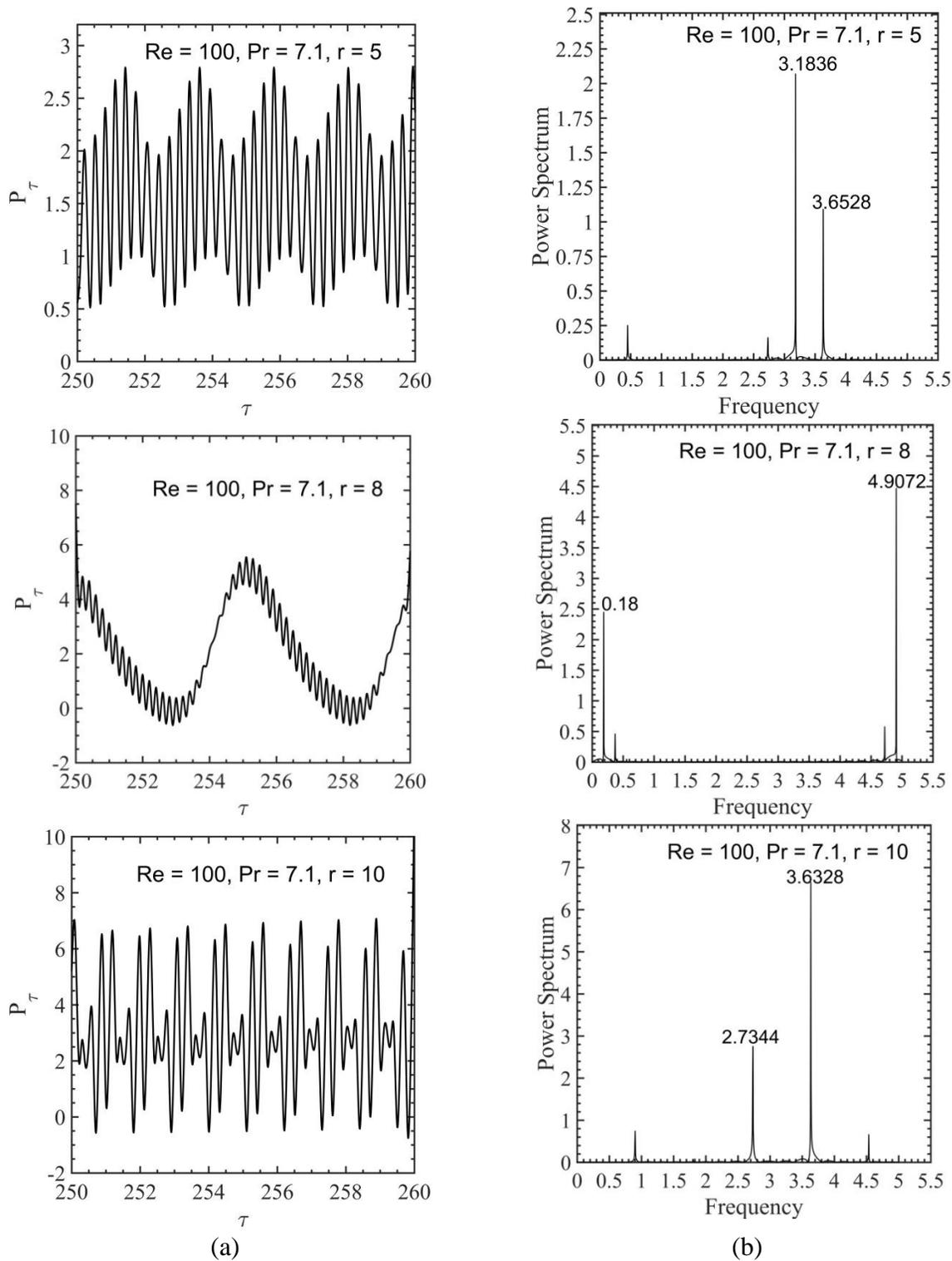

(a) (b)

Figure 13. (a) Temporal variation of inlet pressure, (b) One sided fast Fourier transform power spectrum of inlet pressure for different velocity ratio ($r$ = 5, 8, 10) at $Re$ = 100, $Ri$ = 1, $Pr$ = 7.1

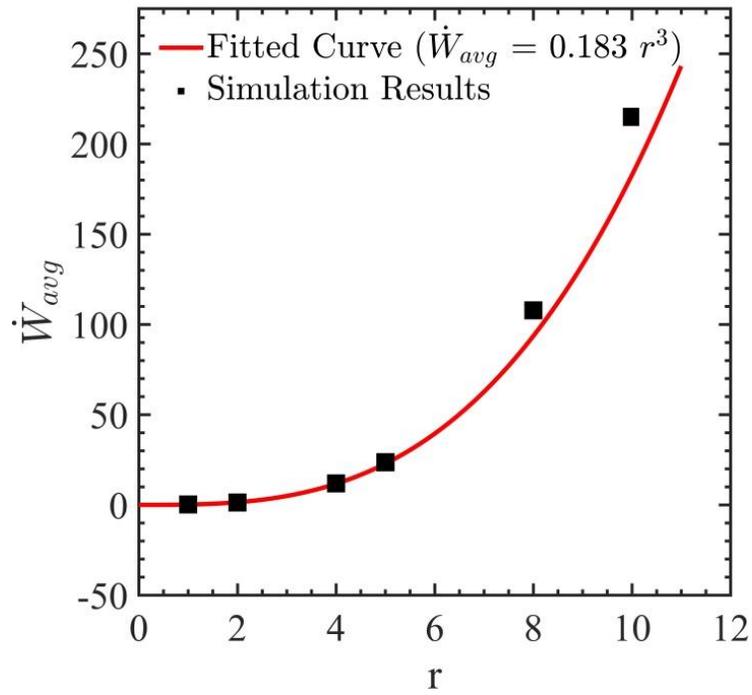

Figure 14. Variation of time average power ($\dot{W}_{avg}$) with velocity ratio (*r*) at *Re* = 100, *Ri* = 1, *Pr* = 7.1